\documentclass[aps,pra,a4paper,twocolumn,showpacs,showkeys]{revtex4}
\usepackage{graphicx,amsmath,bbm,mathrsfs,amssymb}
\newcommand{\ket}[1]{\vert #1 \rangle} \newcommand{\bra}[1]{\langle #1 \vert}
 \newcommand{\sfy}{{\sf y}}

\newcommand{\bmsigma}{\boldsymbol \sigma} \newcommand{\bmSigma}{\boldsymbol \Sigma}
\newcommand{\bmX}{\boldsymbol X} \newcommand{\bmA}{\boldsymbol A}
\newcommand{\bmB}{\boldsymbol B} \newcommand{\bmC}{\boldsymbol C}
\newcommand{\bmS}{\boldsymbol S} \newcommand{\bmLambda}{{\boldsymbol \Lambda}}
 \newcommand{\sfx}{{\sf x}}
\newcommand{\Aop}{{\mathscr A}} 
 \newcommand{\Cop}{{\mathscr C}}
\newcommand{\Xop}{{\mathscr X}} \newcommand{\Fop}{{\mathscr F}}
\newcommand{\Gop}{{\mathscr G}} 
 \newcommand{\oF}{{\overline F}}
\begin{document}
\title{Cloning of Gaussian states by linear optics}
\author{Stefano Olivares} \email{Stefano.Olivares@mi.infn.it}
\author{Matteo G.~A.~Paris}
\affiliation{Dipartimento di Fisica dell'Universit\`a degli Studi di Milano,
Italia.}
\author{Ulrik L.~Andersen} \email{andersen@kerr.physik.uni-erlangen.de}
\affiliation{Institut f\" ur Optik, Information und Photonik, Max-Planck
Forschungsgruppe, Universit\" at Erlangen-N\" urnberg, G\" unther-Scharowsky
str.~1, 91058, Erlangen, Germany}
\begin{abstract}
We analyze in details a scheme for cloning of Gaussian states
based on linear optical components and homodyne detection
recently demonstrated by U.~L.~Andersen {\em et al.}~[Phys.~Rev.~Lett.~{\bf 94}, 
240503 (2005)].  The input-output fidelity is evaluated for a generic (pure or mixed)
Gaussian state taking into account the effect of non-unit quantum
efficiency and unbalanced mode-mixing. In addition, since in most
quantum information protocols the covariance matrix of the set of
input states is not perfectly known, we evaluate the average
cloning fidelity for classes of Gaussian states with the degree
of squeezing and the number of thermal photons being only
partially known. 
\end{abstract}
\date{\today}
\pacs{03.67.Hk, 03.65.Ta, 42.50.Lc}
\keywords{Quantum cloning, Gaussian states, linear optics}
\maketitle
\section{Introduction}\label{s:intro}
The generation of perfect copies of an unknown quantum state is
impossible according to the very nature of quantum mechanics. This
is succinctly formulated by the no-cloning
theorem~\cite{wooters82.nat,dieks82.pla,cl3,cl4}.  It is, however,
possible to make approximate copies of a quantum state by using
a quantum cloning machine~\cite{buzek96.pra}.
Originally, such a machine was proposed for cloning of qubits and
has later been demonstrated experimentally~\cite{dv:experiment}.
Shortly after this development, a continuous variable
(CV)~\cite{braunstein05.rev} analog of the qubit quantum cloner
was proposed~\cite{cl:cerf,cerf:PRA:2000} and recently it was
shown that a CV optimal Gaussian cloner of coherent states can be
implemented using an appropriate combination of beam splitters
and a single phase insensitive parametric
amplifier~\cite{braunstein01.prl,fiurasek01.prl}.  Although this
proposal sounds experimentally promising, the implementation of
an efficient phase insensitive amplifier operating at the
fundamental limit is a challenging task. This problem was solved
by Andersen et al.~\cite{andersen05.prl}, who proposed and
experimentally realized a much simpler configuration for optimal
cloning of coherent states. The realization relies on simple
linear optical components and a feed-forward loop. 
As a consequence of the simplicity, as well as the high quality 
of the optical devices used in this experiment, performances
close to optimal ones were attained. In turn, the resulting 
cloning machine represents a highly versatile tool for further 
investigations on transformation of quantum information from a 
single system to many systems. 
\par
A commonly used figure of merit to quantify the performance of
cloning machines is the fidelity which is a measure of similarity
between the hypothetically perfect clone, {\em i.e} the input
state, and the actual clone.  If the cloning fidelity is
independent on the initial state the machine is referred to as a
{\em universal} cloner.  On the other hand, if the efficiency of
the cloning action depends on the input state, then the proper
measure in order to assess the performances of the machine is the
average fidelity, which weight the fidelities associated to
possible input states with the corresponding occurrence
probability. In other words, for non-universal cloners, the
alphabet of input states, and the distribution thereof, must be
taken into account while evaluating the fidelity. Such an average
fidelity has been considered in
\cite{cochrane04.pra,braunstein00.mod,hammerer05.prl}. However,
in all these references it is assumed that the input alphabet is
only consisting of coherent states, hereby keeping the covariance
matrix of all the possible input states constant. On the other
hand, in some experimental realizations, the covariance matrix is
not perfectly known due to uncontrollable fluctuations, and
therefore it is important to include this uncertainty into the
analysis.
\par
The aim of this paper is two-fold. At first we present a thorough
theoretical description of the cloning machine described in 
Ref.~\cite{andersen05.prl} using a suitable phase-space analysis.
In this way the full quantum dynamics of the machine can be taken
into account; in particular we include the effect of losses in
the detection scheme, as well as variations in the setups beam
splitter ratios. The second topic of the paper is to investigate
the average fidelity of the cloning machine for different
ensembles of input states such as sets made of displaced squeezed 
or displaced thermal states with the squeezing parameter, or the 
number of thermal photons, distributed according to predefined 
distributions. 
\par
The paper is structured as follows: in Sec.~\ref{s:scheme} we
review the main components of the cloning machine based on linear
optics, whereas in Sec.~\ref{s:GScloning} we calculate the
input-output fidelities for the case of generic Gaussian states,
and for specific classes including coherent, displaced squeezed and 
displaced thermal states. Finally, Sec.~\ref{s:outro} closes
the paper with some concluding remarks.
\section{The linear cloning machine}\label{s:scheme}
\begin{figure}[h]
\includegraphics[width=.4\textwidth]{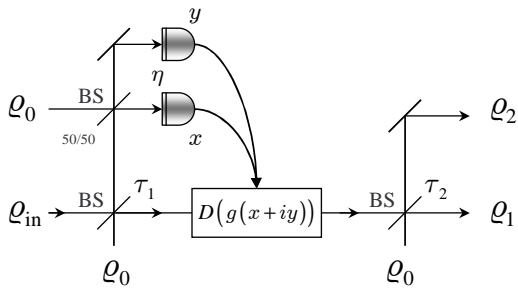}
\vspace{-.2cm}
\caption{\label{f:cl:scheme} Cloning of Gaussian states by linear
optics: the input state $\varrho_{\rm in}$ is mixed with the
vacuum $\varrho_{0}$ at a beam splitter (BS) of 
transmissivity $\tau_{1}$. The reflected beam is measured 
by double-homodyne detection and the outcome of the measurement
$x + i y$ is forwarded to a modulator, which imposes a 
displacement $g (x + iy)$ on the transmitted beam, $g$ being a 
suitable amplification factor. Finally, the displaced state is 
impinged onto a second beam splitter of transmissivity
$\tau_{2}$. The two outputs, $\varrho_1$ and $\varrho_2$, 
from the beam splitter represents the two clones.}
\end{figure}

Optimal Gaussian cloning can be realized using a phase
insensitive amplifier and a beam
splitter~\cite{braunstein01.prl,fiurasek01.prl}.  However, it has
been recently shown, theoretically and experimentally, that the
parametric amplifier can be replaced by a simpler scheme involving
only linear optical components, homodyne detection and a
feed-forward loop~\cite{andersen05.prl}. This scheme, which is
schematically depicted in Fig.~\ref{f:cl:scheme}, will be
referred to as the linear cloning machine throughout the paper. 
\par
The input state, denoted by the density operator $\varrho_{\rm in}$, is 
mixed with the vacuum at a beam splitter (BS) with transmittivity 
$\tau_1$. On the reflected part, double-homodyne detection is performed 
using two detectors with equal quantum efficiencies $\eta$: this 
measurement is executed by splitting the state at a balanced
beam splitter and, then, measuring the two conjugate quadratures $\hat x =
\frac{1}{\sqrt{2}}(\hat{a}+\hat{a}^{\dag})$ and $\hat y =
\frac{1}{i\sqrt{2}}(\hat{a} - \hat{a}^{\dag})$, with $\hat{a}$ and
$\hat{a}^\dagger$ being the field annihilation and creation operator. The
outcome of the double-homodyne detector gives the complex number $\alpha = x + i
y$. According to these outcomes, the transmitted part of the input state
undergoes a displacement by an amount $g \alpha$, where $g$ is a suitable
electronic amplification factor, and, finally, the two output states,
denoted by the density operators $\varrho_1$ and $\varrho_2$, are obtained
by dividing the displaced state using another beam splitter with
transmittivity $\tau_2$.  When $\tau_1 = \tau_2 = 1/2$, $g = 1$ and $\eta =
1$, the scheme reduces to that of Ref.~\cite{andersen05.prl}, which was
shown to be optimal for Gaussian cloning of coherent states on the
basis of a description in the Heisenberg picture. 
Here we apply a different approach which captures all the essential features of the machine.
Towards this aim, in the 
following we carry out a thorough description of the machine using the
characteristic function approach.  
\par
The characteristic function $\chi_{\rm in}(\bmLambda_1) \equiv
\chi[\varrho_{\rm in}](\bmLambda_1)$ associated with a generic Gaussian input
state $\varrho_{\rm in}$ of mode $1$ reads:
\begin{equation}\label{rho:in}
\chi_{\rm in}(\bmLambda_1) = \exp\left\{ -\mbox{$\frac12$} \bmLambda_1^T
\bmsigma_{\rm in}\,\bmLambda_1 - i \bmLambda_1^T \bmX_{\rm in}\right\}\,,
\end{equation}
where $\bmLambda_1 = (\sfx_1, \sfy_1)^T$, $(\cdots)^T$ denotes the
transposition operation, and
\begin{equation}\label{m:covarianza}
\bmsigma_{\rm in} =
\left(
\begin{array}{cc}
\gamma_{11} & \gamma_{12}\\
\gamma_{21} & \gamma_{22}
\end{array}
\right)\,,
\end{equation}
with $\gamma_{12}=\gamma_{21}$, is the covariance matrix.  $\bmX_{\rm in} =
{\rm Tr}[\varrho_{\rm in}\, (\hat x, \hat y)^T]$ is the vector of mean
values, $\hat x$ and $\hat y$ being the quadrature operators defined above.
The vacuum state $\varrho_0 = \ket{0}\bra{0}$ of mode $2$ is described
by the (Gaussian) characteristic function
\begin{equation}
\chi_{0}(\bmLambda_2) \equiv
\chi[\varrho_0](\bmLambda_2) =
\exp\left\{ -\mbox{$ \frac12 $} \bmLambda_2^T \bmsigma_0\, \bmLambda_2
\right\}\,,
\end{equation}
where $\bmsigma_0 = \frac12 \mathbbm{1}_2$,
$\mathbbm{1}_2$ being the $2 \times 2$ identity matrix. In turn, the
initial two-mode state $\varrho = \varrho_{\rm in} \otimes \varrho_0$ is
Gaussian and its two-mode characteristic function reads:
\begin{equation}
\chi[\varrho](\bmLambda)
= \exp\left\{ -\mbox{$\frac12$} \bmLambda^T
\tilde{\bmsigma} \,\bmLambda - i \bmLambda^T \tilde{\bmX} \right\}\,,
\end{equation}
with
\begin{equation}
\tilde{\bmsigma} = \left(
\begin{array}{c|c}
\bmsigma_{\rm in} & {\boldsymbol 0} \\
\hline
{\boldsymbol 0} & \bmsigma_0
\end{array}
\right)\,,\qquad
\tilde{\bmX} = (\bmX_{\rm in}, {\boldsymbol 0})^T\,,
\end{equation}
and $\bmLambda = (\bmLambda_1, \bmLambda_2)^T$.
Under the action of the first BS the state $\chi[\varrho](\bmLambda)$
preserves its Gaussian form, namely
\begin{equation}
\chi[\varrho](\bmLambda) \rightsquigarrow \chi[\varrho'](\bmLambda) 
= \exp\left\{ -\mbox{$\frac12$} \bmLambda^T
\bmsigma \,\bmLambda - i \bmLambda^T \bmX \right\}\,,
\end{equation}
where $\varrho' = U_{{\rm BS},1}\,\varrho_{\rm in}\otimes\varrho_0\,U_{{\rm
BS},1}^{\dag}$, while its covariance matrix and mean values 
transform as~\cite{FOP:napoli:05}:
\begin{align}
&\tilde{\bmsigma} \rightsquigarrow \bmsigma \equiv
{\bmS}_{{\rm BS},1}^T\, \tilde{\bmsigma}\,{\bmS}_{{\rm BS},1} =
\left(
\begin{array}{c|c}
\bmA & \bmC\\
\hline
\bmC^T & \bmB
\end{array}
\right)\,,\label{transf:cvm}\\
&\tilde{\bmX} \rightsquigarrow
\bmX \equiv {\bmS}_{{\rm BS},1}^T\,\tilde{\bmX} = (\bmX_1,\bmX_2)^T\,,
\label{transf:ave}
\end{align}
$\bmA$, $\bmB$, and $\bmC$ are $2 \times 2$ matrices, and
\begin{equation}\label{symp:BS}
\bmS_{{\rm BS},1} =
\left(
\begin{array}{c|c}
\sqrt{\tau_1}\, \mathbbm{1}_2 & \sqrt{1-\tau_1}\, \mathbbm{1}_2 \\
\hline
-\sqrt{1-\tau_1}\, \mathbbm{1}_2  & \sqrt{\tau_1}\, \mathbbm{1}_2
\end{array}
\right)\,,
\end{equation}
is the symplectic transformation associated with the evolution operator
$U_{{\rm BS},1}$ of the BS with transmission $\tau_1$.  
Note that $\varrho'$ is an entangled state if the set of 
states to be cloned consists of nonclassical states, {\em i.e.} 
states with singular Glauber P-function or negative Wigner 
function \cite{visent,wang}.
\par
The subsequent step is to describe double-homodyne detection with quantum
efficiency $\eta$ on the reflected beam. 
This action can be described by the following positive
operator-valued measure (POVM):
\begin{equation}\label{DH:POVM}
\Pi_{\eta}(\alpha) = \int_{\mathbb{C}} d^2\xi\,
\frac{1}{\pi \sigma_\eta}\exp\left\{ -\frac{|\alpha - \xi|^2}
{\sigma_\eta} \right\} \frac{\ket{\xi}\bra{\xi}}{\pi}\,,
\end{equation}
where $\sigma_\eta = (1-\eta)/\eta$ and $\ket{\xi}$ is a coherent state.
Eq.~(\ref{DH:POVM}) describes a Gaussian measurement, the
characteristic function associated with $\Pi_{\eta}(\alpha)$ has the form
\begin{equation}
\chi[\Pi_{\eta}(\alpha)](\bmLambda_2) =\frac{1}{\pi}
\exp\left\{ -\mbox{$\frac12$} \bmLambda_2^T\,\bmsigma_{\rm M}\,\bmLambda_2 -
i \bmLambda_2^T\,\bmX_{\rm M} \right\}\,,
\end{equation}
with $\bmX_{\rm M} = \left({\rm Re}[\alpha],{\rm Im}[\alpha]\right)^T$ and
\begin{equation}\label{sigma:M}
\bmsigma_{\rm M} = \Delta^2\,\mathbbm{1}_2, \qquad \Delta^2 =
\frac12 + \sigma_{\eta} = \frac{2 - \eta}{2\eta}\,.
\end{equation}
The probability of obtaining the outcome $\alpha$ is given by
\begin{align}
p_{\eta}(\alpha) &= {\rm Tr}_{12}[\varrho'\,
\mathbb{I}\otimes\Pi_{\eta}(\alpha)]\\
&= \frac{1}{(2\pi)^2} \int_{\mathbb{R}^4}\!\!\! d^4\bmLambda\,
\chi[\varrho'](\bmLambda)\,
\chi[\mathbb{I}\otimes\Pi_{\eta}(\alpha)](-\bmLambda)\\
&=\frac{
\exp\left\{ -\mbox{$\frac12$}(\bmX_{\rm M}-\bmX_{2})^T\,
\bmSigma^{-1}\,(\bmX_{\rm M}-\bmX_{2})\right\}}
{\pi \sqrt{{\rm Det}[\bmSigma]}}\,,
\end{align}
where $\chi[\mathbb{I}\otimes\Pi_{\eta}(\alpha)](\bmLambda)\equiv
\chi[\mathbb{I}](\bmLambda_1)\,
\chi[\Pi_{\eta}(\alpha)](\bmLambda_2)$, 
$\chi[\mathbb{I}](\bmLambda_1) =
2\pi \delta^{(2)}(\bmLambda_1)$ and $\delta^{(2)}(\zeta)$ is the complex
Dirac's delta function. We also introduced the $2\times 2$ matrix
$\bmSigma = \bmB + \bmsigma_{\rm M}$. 
\par
The conditional state $\varrho_{\rm c}$ of the transmitted beam, obtained when 
the outcome of  the measurement is $\alpha$, i.e.,
\begin{equation}
\varrho_{\rm c} =
\frac{{\rm Tr}_{2}[\varrho'\,\Pi_{\eta}(\alpha)]}{p_{\eta}(\alpha)}\,,
\end{equation}
has the following characteristic function (for the sake of clarity we
explicitly write the dependence on $\bmLambda_1$ and $\bmLambda_2$)
\begin{align}
\chi[\varrho_{\rm c}](\bmLambda_1)
=&
\int_{\mathbb{R}^2}\!\!\! d^2\bmLambda_2\,
\frac{
\chi[\varrho'](\bmLambda_1,\bmLambda_2)\,
\chi[\Pi_{\eta}(\alpha)](-\bmLambda_2)}
{p_{\eta}(\alpha)}\\
=&\exp\left\{
-\mbox{$\frac12$}\bmLambda_1^T
\left[ \bmA - \bmC\bmSigma^{-1}\bmC^T \right]
\bmLambda_1 \right.\nonumber\\
&\left. -\mbox{$\frac12$}\bmX_2^T\,\bmSigma^{-1}\, \bmX_2
+i \bmLambda_1^T \left[ \bmC\bmSigma^{-1}\,\bmX_2 - \bmX_1 \right]
\right\}\nonumber\\
&\times
\exp\left\{ -\mbox{$\frac12$}\bmX_{\rm M}^T\,\bmSigma^{-1}\, \bmX_{\rm M}
\right.\nonumber\\
&\hspace{0.2cm}
\left. +i \bmX_{\rm M}^T \left[ i\bmSigma^{-1}\,\bmX_2 +\bmSigma^{-1}\bmC^T
\bmLambda_1 \right]\right\}\,.
\end{align}
Now, the conditional state $\varrho_{\rm c}$ is displaced by the amount
$g \alpha$ resulting from the measurement amplified by a factor $g$.
By averaging over all possible outcomes of the double-homodyne
detection, we obtain the following output state:
\begin{equation}\label{rho:d}
\varrho_{\rm d} = \int_{\mathbb{C}}d^2 \alpha\,
p_{\eta}(\alpha)\,D(g \alpha)\, \varrho_{c} \,D^{\dag}(g \alpha)\,,
\end{equation}
with $D(z)$ being the displacement operator. In turn, the characteristic
function reads as follows:
\begin{equation}
\chi[\varrho_{\rm d}](\bmLambda_1) = 2\exp\left\{
-\mbox{$\frac12$} \bmLambda_1^T\,\bmsigma_{\rm d}\,\bmLambda_1
-i \bmLambda_1^T \bmX_{\rm d} \right\}\,,
\end{equation}
with $\bmsigma_{\rm d} = \bmA + g(\bmSigma + 2 \bmC^T)$ and
$\bmX_{\rm d} = \bmX_1 + g \bmX_2$.
The conditioned state (\ref{rho:d}) is then sent to a second beam
splitter with transmission $\tau_2$ (see Fig.~\ref{f:cl:scheme}),
where it is mixed with the vacuum $\varrho_0$, and finally the
two clones are generated. Note that, in practice, the average
over all the possible outcomes $\alpha$ in Eq. (\ref{rho:d})
should be performed at this stage, that is after the second beam
splitter. On the other hand, because of the linearity of the
integration, the results are identical, but performing the
averaging just before the beam splitter simplifies the
calculations. Since $\varrho_{\rm d}$ is still Gaussian, the
two-mode state $\varrho_{\rm f} = \varrho_{\rm d} \otimes
\varrho_0$ is a Gaussian with covariance matrix and mean given by
\begin{equation}
\bmsigma_{\rm f} = \left(
\begin{array}{c|c}
\bmsigma_{\rm d} & {\boldsymbol 0} \\
\hline
{\boldsymbol 0} & \bmsigma_0
\end{array}
\right)\,,\qquad
\bmX_{\rm f} = (\bmX_{\rm d}, {\boldsymbol 0})^T\,,
\end{equation}
respectively, which, as in the case of Eqs.~(\ref{transf:cvm}) and
(\ref{transf:ave}), under the action of the BS transform as follows:
\begin{align}
&\bmsigma_{\rm f} \rightsquigarrow \bmsigma_{\rm out} \equiv
{\bmS}_{{\rm BS},2}^T\, \bmsigma_{\rm f}\,{\bmS}_{{\rm BS},2} =
\left(
\begin{array}{c|c}
\Aop_1 & \Cop\\
\hline
\Cop^T & \Aop_2
\end{array}
\right)\,,\label{transf:cvm:fin}\\
&\bmX_{\rm f} \rightsquigarrow
\bmX_{\rm out} \equiv {\bmS}_{{\rm BS},2}^T\,\bmX_{\rm f} =
(\Xop_1,\Xop_2)^T\,,
\label{transf:ave:fin}
\end{align}
where $\Aop_k$ and $\Cop$ are $2 \times 2$ matrices, and $\bmS_{{\rm BS},2}$ is
the symplectic matrix given by Eq.~(\ref{symp:BS}) with $\tau_1$ replaced by 
$\tau_2$. Finally, the (Gaussian) characteristic function of the
clone $\varrho_k$, $k=1,2$, is obtained by integrating over $\bmLambda_{h}$,
$h\ne k$, the two-mode characteristic function $\chi[\varrho_{\rm
out}](\bmLambda_1,\bmLambda_2)$, where $\varrho_{\rm out} = U_{{\rm BS},2}\,
\varrho_{\rm f}\otimes \varrho_0 \,U_{{\rm BS},2}^{\dag}$, i.e.,
\begin{align}
\chi[\varrho_k](\bmLambda_k) &= \frac{1}{2\pi}
\int_{\mathbb{R}^2}\!\!\! d^2\bmLambda_h\,
\chi[\varrho_{\rm out}](\bmLambda_1,\bmLambda_2)\\
&=\exp\left\{ -\mbox{$\frac12$}\bmLambda_{k}^T\,\Aop_k\,\bmLambda_{k}
- i \bmLambda_k^T\, \Xop_k\right\}\,.\label{clone:k}
\end{align}
Let us now focus our attention on $\bmX_{\rm out}$: the explicit
expressions of $\Xop_1$ and $\Xop_2$ are
\begin{align}
&\Xop_1 = \sqrt{\tau_2}\left(  \sqrt{\tau_1} + g\sqrt{1-\tau_1}\right)
\bmX_{\rm in}\,,\\
&\Xop_2 = \sqrt{1-\tau_2}\left(  \sqrt{\tau_1} + g\sqrt{1-\tau_1}\right)
\bmX_{\rm in}\,.
\end{align}
As a matter of fact, in order to have two output Gaussian states with the
same means $\Xop_1 = \Xop_2$, one should put $\tau_2 = 1/2$; furthermore,
if one also sets
\begin{equation}\label{g:symm}
g = g_{\rm s} \equiv \sqrt{\frac{2}{1-\tau_1}}-
\sqrt{\frac{\tau_1}{1-\tau_1}}
\,,
\end{equation}
then $\Xop_1=\Xop_2=\bmX_{\rm in}$, corresponding to unity gain cloning. On
the other hand, $\Aop_{k}$ can be written in a compact form as follows:
\begin{subequations}\label{A:k}
\begin{align}\Aop_{1} &= \mbox{$\frac12$} (1-\tau_2)\,\mathbbm{1} +
\tau_2\, \mathbf{\Gamma}(\bmsigma_{\rm in})\,,\\
\Aop_{2} &= \mbox{$\frac12$} \tau_2\,\mathbbm{1} +
(1-\tau_2)\, \mathbf{\Gamma}(\bmsigma_{\rm in})\,,
\end{align}
\end{subequations}
where
\begin{equation}
\mathbf{\Gamma}(\bmsigma_{\rm in}) =
\left(
\begin{array}{cc}
\Fop(\gamma_{11}) & \Gop(\gamma_{12})\\
\Gop(\gamma_{21}) & \Fop(\gamma_{22})
\end{array}
\right)\,,
\end{equation}
with
\begin{align}
&\Fop(\gamma) = 1 - \tau_1 +
g\left[\tau_1 - 2 \sqrt{(1-\tau_1)\tau_1}+\Delta^2\right] + \Gop(\gamma)\,,\\
&\Gop(\gamma) = \left[\tau_1 + g\left(1 - \tau_1
+ 2\sqrt{(1-\tau_1)\tau_1}\right)\right]\gamma\,.
\end{align}
Now, if $\tau_2=1/2$ and $g=g_{\rm s}$, one has $\Aop_1 = \Aop_2$ and $\Xop_1=\Xop_2$, as we have
seen above, i.e., the cloning becomes {\em symmetric}. Furthermore, when
also $\tau_1 = 1/2$, thanks to Eqs.~(\ref{clone:k}) and (\ref{A:k}) we have
that the cloning map for the scheme in Fig.~\ref{f:cl:scheme} is given by
the following Gaussian map:
\begin{equation}\label{cl:map}
{\cal G}_{\sigma_{\rm GN}}(\varrho_{\rm in}) = \int_{\mathbb{C}}
\frac{d^2\gamma}{\pi \sigma_{\rm GN}^2}\,
\exp\left\{ - \frac{|\gamma|}{\sigma_{\rm GN}^2} \right\}\,
D(\gamma)\,\varrho_{\rm in}\,D^{\dag}(\gamma)\,,
\end{equation}
where $\sigma_{\rm GN}^2 = \frac12 + \Delta^2$.
Finally, although
$\Cop$ does not appear in Eq.~(\ref{clone:k}), for the sake of
completeness, we give its analytic expression:
\begin{equation}
\Cop = \sqrt{(1-\tau_2)\tau_2}\left[ \mbox{$\frac12$}\mathbbm{1}
-\mathbf{\Gamma}(\bmsigma_{\rm in})\right]\,.
\end{equation}
\par
In the following we will analyze the input-output fidelities for a generic
(pure or mixed) Gaussian state. In particular, we will consider three
classes of Gaussian states, i.e. coherent, displaced squeezed and displaced thermal
states.
\section{Cloning of Gaussian states}\label{s:GScloning}
\subsection{Fidelity}
Usually, the performance of cloning machines are quantified by
the fidelity which is a measure of the similarity between the
hypothetically perfect clone and the actual clone. In its most
general form, the fidelity is given by the Uhlmann's transition
probability
\cite{uhlman:RepMP:76}
\begin{equation}\label{fidelity}
{F}(\varrho_{\rm in},\varrho_k) = \left(
{\rm Tr}\left[
\sqrt{\sqrt{\varrho_{\rm
in}}\,\varrho_{k}\,\sqrt{\varrho_{\rm in}}}
\right]
\right)^2\,,
\end{equation}
and satisfies the natural axioms
\begin{itemize}
\item ${F}(\varrho_{\rm in},\varrho_k)\le 1$ and
${F}(\varrho_{\rm in},\varrho_k) = 1$ if and only if
$\varrho_{\rm in} = \varrho_k$;
\item ${F}(\varrho_{\rm in},\varrho_k) =
{F}(\varrho_k,\varrho_{\rm in})$;
\item if $\varrho_{\rm in}$ is a pure state $\varrho_{\rm in} =
\ket{\psi_{\rm in}}\bra{\psi_{\rm in}}$, then we have
${F}(\varrho_{\rm in},\varrho_k) =
\bra{\psi_{\rm in}} \varrho_{k} \ket{\psi_{\rm in}}$;
\item ${F}(\varrho_{\rm in},\varrho_k)$ is invariant under unitary
transformations on the state space.
\end{itemize}  
Furthermore, when $\varrho_{\rm in}$ and $\varrho_k$ are Gaussian states of
the form (\ref{rho:in}) and (\ref{clone:k}), the fidelity (\ref{fidelity})
becomes \cite{scutaru:JPA:98,nha:2005} 
\begin{align}\label{gen:fid}
&{F}_{\eta} \equiv {F}(\varrho_{\rm in},\varrho_k) =
\frac{1}
{
\sqrt{{\rm Det}[\bmsigma_{\rm in}+\Aop_k]+\delta}-\sqrt{\delta}
}\nonumber\\
&\times\exp\left\{
-\mbox{$\frac12$} (\bmX_{\rm in}-\Xop_k)^T(\bmsigma_{\rm in}+\Aop_{k})^{-1}
(\bmX_{\rm in}-\Xop_k)
\right\}
\,,
\end{align}
where $\delta = 4({\rm Det}[\bmsigma_{\rm in}]-\frac14)
({\rm Det}[\Aop_k]-\frac14)$. Note that for pure Gaussian
states ${\rm Det}[\bmsigma_{\rm in}] = \frac14$, and in turn $\delta = 0$.
\par
For {\em symmetric} cloning, i.e. for $\tau_2 = 1/2$ and $g = g_{\rm
s}$ in Eq.~(\ref{g:symm}), then Eq.~(\ref{gen:fid}) reduces to
\begin{equation}\label{gen:fid:symm}
{F}_{\eta} =
\frac{1}
{\sqrt{{\rm Det}[\bmsigma_{\rm in}+\Aop_k]+\delta}-\sqrt{\delta}}\,.
\end{equation}
\par
In general, the cloning fidelity in (\ref{gen:fid}) is state
dependent, and therefore the figure of merit to be considered is 
the mean cloning fidelity, averaged over the ensemble of possible 
input states. In order to evaluate this quantity, 
we parametrize the input ensemble (class) $\{
\varrho_{\rm in}(\boldsymbol\lambda) \}$ of different Gaussian states,
by $\boldsymbol\lambda\in\Omega$ and consider each of them occurring with 
the {\em a priori} probability $p(\boldsymbol\lambda)$. The average 
fidelity then reads
\begin{equation}\label{ave_fid}
\oF_{\eta} = \int_{\Omega} d\boldsymbol\lambda\,
p(\boldsymbol\lambda)\,{F}_{\eta}(\boldsymbol\lambda)\,.
\end{equation}
Within the set of possible states both the mean values as well as the
covariance matrices may vary. Assuming that the probability distribution
$p(\boldsymbol{\lambda})$ is factorisable into a distribution for the mean values
$p(\alpha)$ and a distribution for the covariance matrix,
$p(\boldsymbol\sigma_{\rm in})$, we may write
$p(\boldsymbol{\lambda})=p(\alpha)\,p(\boldsymbol\sigma_{\rm in})$, and the average
fidelity reads
\begin{equation}
\oF_{\eta} = \int_{\Omega} d\boldsymbol\sigma_{\rm in}\,d\alpha\,
p(\alpha)\,p(\boldsymbol\sigma_{\rm in})\,
{F}_{\eta}(\boldsymbol\sigma_{\rm in},\alpha)\,.
\end{equation}
In the extreme case where both $\bmsigma_{\rm in}$ and $\alpha$ are fixed,
the input state is completely known and perfect cloning with unit fidelity
is of course possible. A more interesting scenario is when the covariance
matrix is fixed, as for example the case in which the set is made
by coherent states, while the displacement (that is, the mean value) 
is random. In this case the average fidelity reduces to 
\begin{equation}\label{F:alpha}
\oF_{\eta} = \int_{\mathbb C} d^2\alpha\,p(\alpha)\,
{F}_{\eta}(\alpha)
\end{equation}
If $\tau_1=1/2$ and $g=g_s$, the map (\ref{cl:map}) is covariant with
respect to displacements, meaning that if two input states are identical up
to a displacement their respective clones should be identical up to the
same displacement \cite{cerf:EPJ:2002}. Indeed, if the input state is of
the form $\varrho_{\rm in}(\alpha) = D(\alpha)\,\varrho_{\rm
s}\,D^{\dag}(\alpha)$, $\varrho_{\rm s}$ being a seed state, then the
fidelity ${F}_{\eta}(\alpha)$ actually does not depend on complex parameter
$\alpha$, and, as consequence, $\oF_{\eta} = {F}_{\eta}$.  
Therefore, in this case the noise added by the cloning process
(\ref{cl:map}) is the same noise added in cloning of coherent states, i.e.,
the cloning is {\em optimal}. Notice that the corresponding optimal fidelity
$\oF$ is not necessarily equal to $2/3$ [see Eq.~(\ref{gen:fid:symm})].
\subsection{Coherent states}
\begin{figure}[h]
\vspace{-1cm}
\setlength{\unitlength}{1mm}
\begin{center}
\begin{picture}(70,50)(0,0)
\put(4,0){\includegraphics[width=60\unitlength]{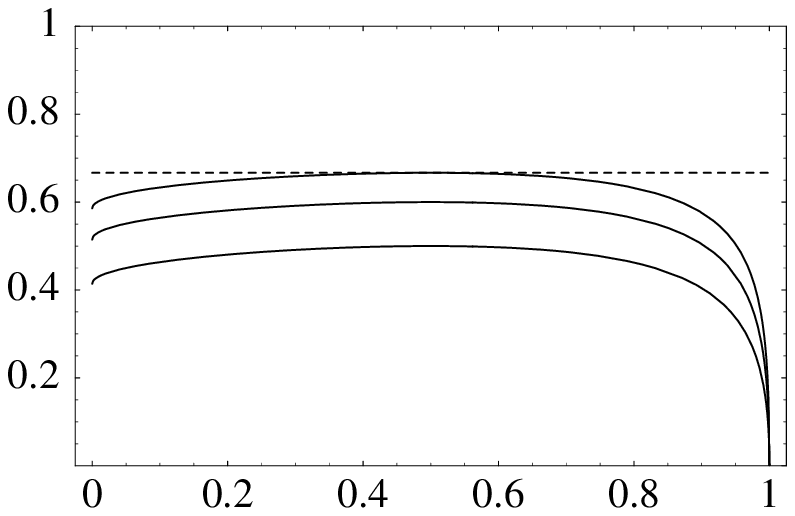}}
\put(35,-3){$\tau_1$}
\put(-2,24.5){$F_{\eta}$}
\end{picture}
\end{center}
\caption{
Linear cloning fidelity ${F}_{\eta}$ for a coherent state
as a function of the BS transmittivity $\tau_1$ for different values of
the quantum efficiency $\eta$: from top to bottom $\eta = 1.0$, $0.75$,
and $0.5$. We set $\tau_2 = 1/2$ and $g = g_{\rm s}$ (symmetric cloning).
The dashed line is the value $2/3$. The fidelity does not depend on the
coherent state amplitude.} \label{f:CS}
\end{figure}

Before addressing the general case let us reconsider 
cloning of (pure) coherent states. For this set of states
our linear machine provides universal cloning, 
{\em i.e.} state independent fidelity.
In Fig.~\ref{f:CS} we plot the fidelity as given by 
Eq.~(\ref{gen:fid:symm}), as a function of $\tau_1$, 
for different values of $\eta$ and for $\tau_2 = 1/2$, 
$g = g_{\rm s}$. In this case, corresponding to symmetric cloning, 
the machine yields the optimal fidelity 
$F=2/3$ predicted for universal Gaussian cloning of coherent states.  
Notice that the optimal fidelity is achieved with $\tau_1 = 1/2$ and 
$\eta = 1$; by expanding the fidelity up to the second order 
around $\tau_1=1/2$  we obtain
\begin{align}
\overline{F}_{\eta} \equiv {F}_{\eta} \simeq  \frac{2\eta}{1+2\eta}\left[
1- \frac{\left(\tau_1-\mbox{$\frac12$}\right)^2}{1+2\eta}
\right] \nonumber \end{align}
>From this expression we clearly see that the cloning machine
proposed in \cite{andersen05.prl} is robust against fluctuations of
the BS ratio. This conclusion can be also directly drawn from
Fig.~\ref{f:CS}.
\par
Let us however note that the fidelity value $F=2/3$ is the optimal one 
only if the input distribution of coherent states is flat, that is, if 
we have no {\em a priori} information about the amplitudes.  If the set 
of coherent states is restricted such that the distribution of amplitudes 
is the Gaussian 
\begin{equation}
p_{\rm a}(\alpha) =\frac{1}{\pi\sigma_{\rm a}^2}
\exp\left\{-\frac{|\alpha|^2}{\sigma_{\rm a}^2}\right\}\,,
\end{equation}
the average fidelity can be increased by choosing a
different gain~\cite{cochrane04.pra}. However, in this scenario the cloning
action becomes state dependent, and the integration in (\ref{F:alpha}) 
should be explicitly performed.
By optimizing the gain we find~\cite{cochrane04.pra}
\begin{align}
\oF& = \frac{2(1+\sigma_{\rm a}^2)}{1+3\sigma_{\rm a}^2},
&{\rm if} \quad \sigma_{\rm a}^2 \geq 1+\sqrt{2},\\
\oF& = \frac{2}{2+(3-2\sqrt{2})\sigma_{\rm a}^2},
&{\rm if} \quad \sigma_{\rm a}^2 < 1+\sqrt{2}\:.
\end{align}
\par
We have now seen that by fixing the covariance matrix of the input states
to coherent states, the fidelity is a function of the distribution (being
delta, flat or Gaussian) of these states. This aspect has been investigated
in the literature~\cite{cochrane04.pra}. In contrast, the case where the
covariance matrix may fluctuate has not received much attention
heretofore.  In the following sections we therefore discuss the average
cloning fidelity for classes of states with covariance matrix 
randomly distributed according to a predetermined distribution. 
We assume that the
displacement of the input state is random and that the cloner is set to
unity gain (that is invariant with respect to the displacement corresponding
to $g=g_s$). In this case the average over the mean value is
trivial and the average fidelity can be written as  
\begin{equation}
\oF_{\eta} = \int_{\Sigma} d\bmsigma_{\rm in}\,
p(\boldsymbol\sigma_{\rm in})\,
{F}_{\eta}(\boldsymbol\sigma_{\rm in})\,.
\end{equation} 
\subsection{Squeezed states}
When the input Gaussian state is the squeezed state $\ket{\alpha,\xi} =
D(\alpha)S(\xi)\ket{0}$, where $D(\alpha) = \exp\{\alpha a^{\dag} -
\alpha^* a\}$ and  $S(\xi) = \exp\{\frac12 (\xi{a^{\dag}}^2 - \xi^* a^2)\}$
are the displacement and squeezing operator, respectively, the entries of
the input covariance matrix (\ref{m:covarianza}) are
\begin{subequations}
\begin{align}
&\gamma_{11} = \mbox{$\frac12$}\left(\cosh 2|\xi|
+ \sinh 2|\xi|\,\cos \varphi\right)\,,\\
&\gamma_{22} = \mbox{$\frac12$}\left(\cosh 2|\xi|
- \sinh 2|\xi|\,\cos \varphi\right)\,,\\
&\gamma_{12} = \gamma_{21} = - \mbox{$\frac12$} \sinh 2|\xi|\,\sin \varphi\,,
\end{align}
\end{subequations}
where we put $\xi = |\xi|\,e^{i\varphi}$; obviously, when $\xi=0$ the
squeezed state $\ket{\alpha,\xi}$ reduces to the coherent state
$\ket{\alpha}$ and $\bmsigma_{\rm in} = \frac12 \mathbbm{1}_2$.  Note that
in this Section we are addressing the case of an {\em unknown} squeezing
parameter $\xi$ (randomly distributed according to a given probability
density): when it is known, the optimal strategy in the Gaussian regime is
to perform the unsqueezing operation $S^{-1}(\xi)$ just before the cloning
machine, proceed as in the case of coherent states, and, at the output
stage, apply the squeezing operation $S(\xi)$ to both the clones which
yields a fidelity of $2/3$ (independent on the amount of fixed squeezing) as
in the coherent state case~\cite{braunstein01.prl}.
\par
\begin{figure}
\vspace{-1cm}
\setlength{\unitlength}{1mm}
\begin{center}
\begin{picture}(70,50)(0,0)
\put(4,0){\includegraphics[width=60\unitlength]{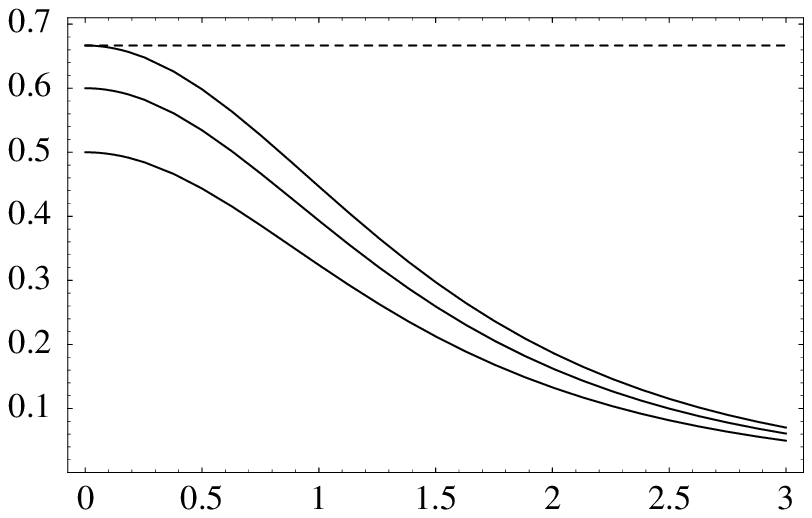}}
\put(35,-3){$|\xi|$}
\put(-2,24.5){$F_{\eta}$}
\end{picture}
\end{center}
\caption{
Plot of the fidelity ${F}_{\eta}(\xi)$ of the squeezed state
$\ket{\alpha, \xi}$ as a function of $|\xi|$ for different values of
$\eta$: from top to bottom $\eta = 1.0$, $0.75$,
and $0.5$. We set $\tau_1 = \tau_2 = 1/2$ and $g = g_{\rm s}$
(symmetric cloning). The dashed line is the value $2/3$.
The fidelity does not depend on the displacement amplitude $\alpha$.}
\label{f:SQfid}
\end{figure}

However in the case of an unknown squeezing parameter the squeezing action
$S(\xi)$ is not known. Therefore in the following, we investigate the
cloning of unknown squeezed states using the cloning machine outlined in
this paper. First we note that since the linear elements involved in the
cloning machine do not affect the phase of the input state, the fidelity
${F}_{\eta}(\xi)$ depends only on $|\xi|$ and, without loss of
generality, we may take $\xi$ as real. The fidelity for Gaussian squeezed
input states, using the coherent state cloning machine, is given by 
\begin{equation}\label{F:r:eta:1:2}
{F}_{\eta}(\xi) =
\frac{4}{\sqrt{(5+2\Delta^2)^2+16(1+2\Delta^2)\sinh^2 |\xi|}}\,.
\end{equation}
This fidelity is plotted in Fig.~\ref{f:SQfid} as a function of the squeezing
parameter for different values of $\eta$. We clearly see that for
coherent states (corresponding to $\xi$=0), the fidelity is $2/3$ while
decreasing with the degree of squeezing, eventually reaching zero for
highly squeezed input states.
\par
In order to calculate the average fidelity, we assume that the squeezed
state $\ket{\alpha,\xi}$ is drawn from an ensemble of states with {\em a
priori} probability $p(\alpha, \xi)= p(\alpha)\,p(\xi)$. Above we mentioned
that the cloning action with unity gain is independent on the distribution
$p(\alpha)$, which can then be left undefined. The distribution of the
squeezing factor is however quite important:  it is clear that for
completely unknown input squeezing (corresponding to a flat distribution)
the average fidelity goes to zero. We therefore must restrict the set of
input squeezed states to, say, a Gaussian distribution given by
\begin{equation}\label{distribution}
p_{\rm s}(\xi) =\frac{1}{\pi\sigma_{\rm s}^2}
\exp\left\{-\frac{|\xi|^2}{\sigma_{\rm s}^2}\right\}\,.
\end{equation}
As evident from this expression we assume the distribution to be centered
at $\xi=0$ which corresponds to a coherent state. This means that the
coherent state is the most likely member in the set of input states, and we
therefore conjecture that our machine is optimal in the Gaussian scenario.
If however the distribution is centered at a known squeezing
amplitude, say $\xi=\xi_0$, then we believe that the optimal machine is the one
mentioned above where the input states are unsqueezed [$S^{-1}(\xi_0)$]
before the cloning machine and squeezed [$S(\xi_0)$] again after the
cloning action. 
\par
\begin{figure}
\vspace{-1cm}
\setlength{\unitlength}{1mm}
\begin{center}
\begin{picture}(70,50)(0,0)
\put(4,0){\includegraphics[width=60\unitlength]{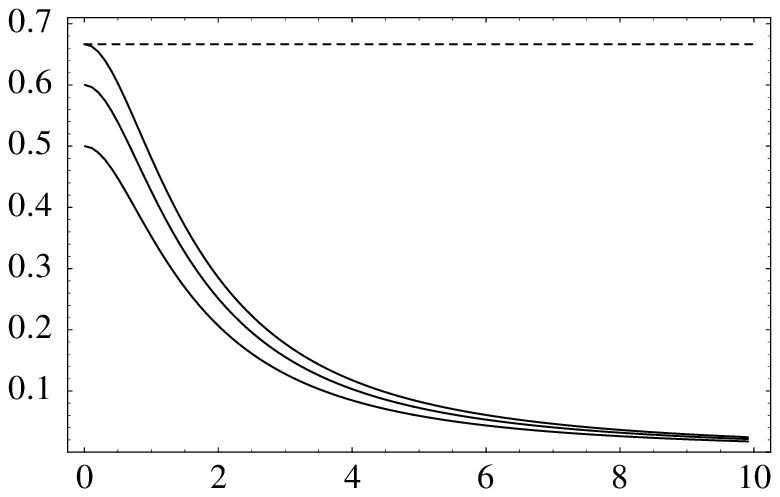}}
\put(35,-3){$\sigma_{\rm s}$}
\put(-2,24.5){$\oF_{\eta}$}
\end{picture}
\end{center}
\vspace{0cm}
\caption{Plot of the average fidelity $\oF_{\eta}$
of a set of squeezed states as a function of $\sigma_{\rm s}$ (see text for
details) and different values of the efficiency $\eta$: form top to bottom
$\eta = 1.0$, $0.75$, and $0.5$.  The dashed line corresponds to $2/3$, i.e.,
the optimal cloning fidelity of coherent states. We put $\tau_1 = \tau_2 =
1/2$ and $g = g_{\rm s}$.} \label{f:ave:squeezed}
\end{figure}

Using the polar coordinates, $d^2\xi =
\rho\, d\rho\, d\phi$, $\xi = \rho\, e^{i\phi}$, and ${F}_{\eta}(\xi) =
{F}_{\eta}(|\xi|)$, the average fidelity now reads
\begin{align}
\oF_{\eta} &= \int_{\mathbb{C}} d^2\xi\,p(\xi)\,{F}_{\eta}(\xi)\\
&= 2 \int_{0}^{+\infty}\!\!\! d\rho\,
\frac{\rho}{\sigma_{\rm s}^2}
\exp\left\{-\frac{\rho^2}{2\sigma_{\rm s}^2}\right\}
\,{F}_{\eta}(\rho) \label{ave:fid:squeezed}
\end{align}
This function is depicted in Fig.~\ref{f:ave:squeezed} as a function of
$\sigma_{\rm s}$ for different values of $\eta$. If the standard deviation
$\sigma _s=0$, the distribution in (\ref{distribution}) is a delta function
and the input alphabet contains only coherent states. In this case it
reduces to the case discussed in the previous section and the expected
fidelity is 2/3 (for ideal detection efficiency) as seen in the figure. We
also see that the fidelity degrades as the width of the distribution of the
squeezing parameter increases, and eventually reaches zero when the a
priori information is poor. At this point we should note that if one allows for non-Gaussian output clones the fidelity can be improved. E.g. it is known that the optimal cloner of coherent states and the optimal universal cloner employ non-Gaussian operations and they yield fidelities of 68.3\%~\cite{cerf05.prl} and 50\%~\cite{braunstein01.pra} respectively. 
\subsection{Thermal states}
\begin{figure}
\vspace{-1cm}
\setlength{\unitlength}{1mm}
\begin{center}
\begin{picture}(70,50)(0,0)
\put(4,0){\includegraphics[width=60\unitlength]{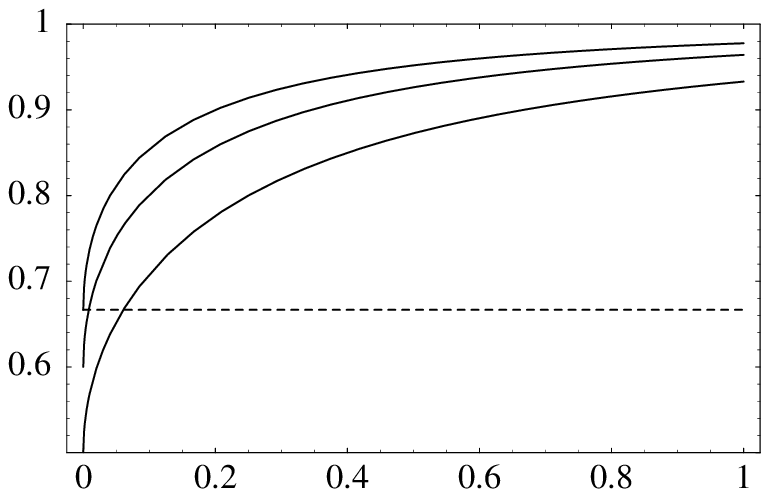}}
\put(35,-3){$N$}
\put(-1,24.5){$F_{\eta}$}
\end{picture}
\end{center}
\vspace{0cm}
\caption{
Plot of the input-output fidelity $F_{\eta}(N)$ of the
displaced thermal state $\varrho_{{\rm th}, \alpha}$ as a function of the
average number of thermal photons $N$ and different values of the
efficiency $\eta$: form top to bottom $\eta = 1.0$, $0.75$, and $0.5$.
The dashed line corresponds to $2/3$, i.e.,
the optimal cloning of coherent states. We put $\tau_1 = \tau_2 = 1/2$ and
$g = g_{\rm s}$.}
\label{f:thermal:2D}
\end{figure}

Another interesting class of Gaussian states is the set of displaced
thermal states $\varrho_{{\rm th},\alpha} = D(\alpha)\,\nu_{\rm th}\,
D^{\dag}(\alpha)$, which arise, for example, from the propagation of
coherent states in a noisy environment \cite{ComEnt}.
The thermal state $\nu_{\rm th}$ is given by 
\begin{equation}
\nu_{\rm th} = \frac{1}{1+N}\, \sum_{m=0}^{\infty}
\left(\frac{N}{1+N}\right)^m \ket{m}\bra{m}\,,
\end{equation}
where $N$ is the average number of thermal photons. Its covariance matrix
is given by $\bmsigma_{\rm in} = (N+\frac12) \mathbbm{1}_2$. Since
$\nu_{\rm th}$ and, in turn, $D(\alpha)\,\nu_{\rm th}\,D^{\dag}(\alpha)$
are not pure states, the cloning fidelity $F_{\rm \eta}(N)$ should be
calculated using the full expression of Eq.~(\ref{gen:fid}), and the
result is plotted in Fig.~\ref{f:thermal:2D} as a function of $N$ and
different values of $\eta$. For the unity gain cloner and
assuming the detection efficiency to be ideal ($\eta =1$), we derive the
expression
\begin{align}
{F}_{\eta=1}(N)=&\bigg(
\frac32 + N (3 + 2 N) \nonumber\\
&-\sqrt{N(2N+1)(2N^2+5N+3)}
\bigg)^{-1}\,.
\end{align} 
We see that the fidelity increases with the average number of thermal
photons, that is, using the fidelity as a measure, the quality of the
cloning action increases with the mixedness of the input states. 
\par
\begin{figure}
\vspace{-1cm}
\setlength{\unitlength}{1mm}
\begin{center}
\begin{picture}(70,50)(0,0)
\put(4,0){\includegraphics[width=60\unitlength]{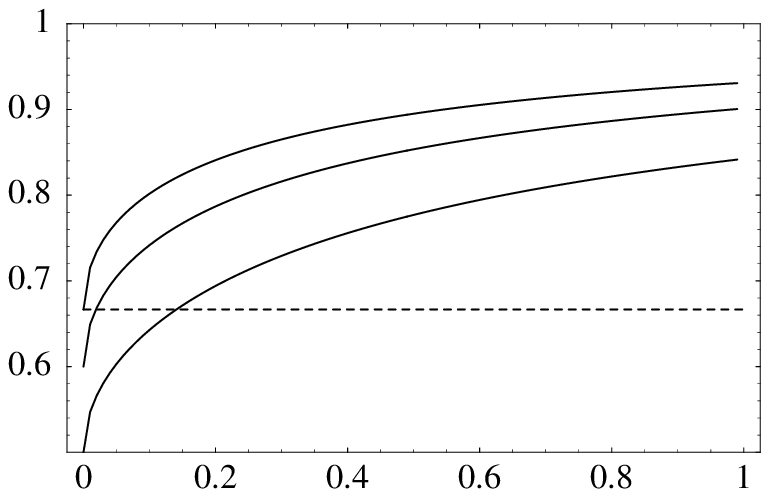}}
\put(35,-3){${\cal N}$}
\put(-2,24.5){$\oF_{\eta}$}
\end{picture}
\end{center}
\vspace{0cm}
\caption{
Plot of the average fidelity $\oF_{\eta}$ of the set of thermal
states distributed according to the top-hat distribution (\ref{top:hat}) as
a function of the threshold value ${\cal N}$ and different values of the
efficiency $\eta$: form top to bottom $\eta = 1.0$, $0.75$, and $0.5$. The
dashed line corresponds to $2/3$. We put $\tau_1 = \tau_2 = 1/2$ and $g =
g_{\rm s}$.}
\label{f:alphabet}
\end{figure}
\begin{figure}
\vspace{-1cm}
\setlength{\unitlength}{1mm}
\begin{center}
\begin{picture}(70,50)(0,0)
\put(4,0){\includegraphics[width=60\unitlength]{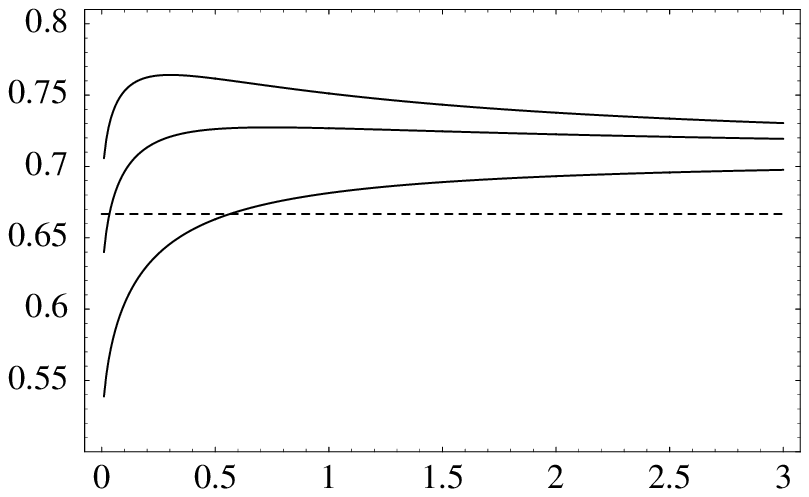}}
\put(35,-3){$\mu_N$}
\put(-2,24.5){$\oF_{\eta}$}
\end{picture}
\end{center}
\vspace{0cm}
\caption{
Plot of the average fidelity $\oF_{\eta}$ of the set of thermal
states distributed according to a ``half-Gaussian'' distribution
(\ref{half:gauss}) as a function of $\mu_N$ and different
values of the efficiency $\eta$: form top to bottom $\eta = 1.0$, $0.75$, and
$0.5$. The dashed line corresponds to $2/3$. We put $\tau_1 = \tau_2 = 1/2$
and $g = g_{\rm s}$.}
\label{f:gauss:alphabet}
\end{figure}
\par
Let us now consider a different ensemble of displaced thermal states, with
random displacement and average number of thermal photons $N$ distributed
around zero either as a bounded flat, top-hat, distribution  or as a 
``half-Gaussian'' distribution. The average fidelity is
\begin{equation}
\oF_{\eta} =
\int_{0}^{+\infty} \!\!\!\! d N\, p(N)\, {F}_{\eta}(N)\,,
\end{equation}
where
\begin{equation}\label{top:hat}
p(N) =
\left\{
\begin{array}{ll}
{\cal N}^{-1} & {\rm if} \quad N \in [0,{\cal N}] \\
0 & {\rm otherwise}
\end{array}
\right.
\end{equation}
for a top-hat distribution, and
\begin{equation}\label{half:gauss}
p(N) =\frac{2}{\sqrt{2\pi \mu_N^2}}
\exp\left\{-\frac{N^2}{2 \mu_N^2}\right\}\,, \qquad (N\geq 0)
\end{equation}
for a (re-normalized) ``half-Gaussian'' distribution.
In Figs.~\ref{f:alphabet} and \ref{f:gauss:alphabet} we show the 
corresponding average fidelities, as functions of ${\cal N}$ and 
$\mu_N$, respectively, for different values of $\eta$.
For the top-hat distribution the average fidelity monotonically
increases as the threshold value ${\cal N}$ increases, whereas
for the half-Gaussian one the average fidelity shows a maximum value
depending on the value of $\eta$, as far as $\eta\gtrsim 0.7$.
\section{Conclusions}\label{s:outro}
We have analyzed in details a recently demonstrated scheme for
linear cloning of Gaussian states \cite{andersen05.prl}. Using a
suitable phase-space analysis the input-output fidelity has been
evaluated for a generic (pure or mixed) Gaussian state taking
into account the effect of non-unit quantum efficiency of homodyne 
detection and fluctuations in the beam splitters transmittivity. 
Our results indicate that the linear cloning
machine suggested in  \cite{andersen05.prl} is robust against
fluctuations of transmissivity and non-unit quantum efficiency.
\par
We have explicitly evaluated the cloning fidelity for specific
classes of non coherent displaced states. We found that a fixed 
(unknown) squeezing of the input states degrades the fidelity with 
respect to the coherent level, as one may expect for 
cloning of highly nonclassical states, while, on the contrary,  
cloning of displaced thermal states may be achieved with larger 
fidelity. Using the above results we have evaluated the average 
cloning fidelity for classes of Gaussian states with fluctuating 
covariance matrix, as for example displaced squeezed or 
displaced thermal states with the degree of squeezing or 
the number of thermal photons randomly distributed according 
to a Gaussian or a uniform distribution.
Results indicate that the average fidelity monotonically 
decreases as the squeezing dispersion increases, whereas 
the behaviour with respect to dispersion of thermal photons
is not monotone.
\section*{Acknowledgments}
Fruitful discussions with A.~Ferraro are kindly acknowledged.
This work has been supported by MIUR through the project PRIN-2005024254-002 and by the EU project COVAQIAL no. FP6-511004.

\end{document}